\documentclass{article}
\usepackage{amsmath}
\usepackage{chicago}

\newtheorem{theorem}{Theorem}
\newtheorem{acknowledgement}[theorem]{Acknowledgement}

\input{tcilatex}

\begin{document}

\title{Rediscussion on gas-liquid phase transition}
\author{Yuan-Xing Gui \\
Department of Physics, Dalian University of Technology, \\
Dalian, 116024, P. R. China}
\maketitle
\date{\today }

\begin{abstract}
Liquid-gas phase transition in statistical mechanics is a long-standing
dilemma not yet well explained. In this paper we propose a novel approach to
this dilemma, by: 1). Putting forth a new space homogeneity assumption. 2).
Giving a new formulation-- the mean distance expansion, instead of Mayer's
Cluster Expansion, to calculate the intermolecular potential and partition
function for a classical system. 3). Explaining how the separation of two
phases occurs below the critical temperature $T_{c}$ and what is the gap
between two phases. 4). Calculating the physical quantities in a system of
coexistent vapor and liquid, and comparing them with the experimental
results. Qualitative, and some quantitative, consistencies are obtained. So
the statistical explanation on first order liquid-gas phase transition is
solved in principle.

\end{abstract}

Gas-liquid phase transition is a fundamental problem remaining unsolved for
centuries in statistical mechanics since the VDW equation of state was
developed in 1873. Since then, a great deal of attention had been focused on
providing statistical explanations for the separation of phases by many
physicists[1-17]. This letter provides a new way to solve it. Based on space
homogeneity assumption and mean distance expansion, instead of Mayer's
Cluster Expansion, to calculate the intermolecular potential and partition
function for a classical system, this work at first time reveals the
existence of the energy gap between two phases; offers a rational physical
explanation on phase transition and latent heat; and demonstrates how to
calculate physical quantities in a system of coexistent vapor and liquid:
such as densities of coexistent gas and liquid, vapor pressure and latent
heat, $etc$. As examples, we discussed gas-liquid transition for both the
VDW fluid and the modified VDW fluid. The experimental results provide
qualitative, and some quantitative supports to the new theory.

H.Eyring and D.Henderson $\mathit{etc}$, mentioned their space homogeneity
assumption (it is referred as EHSE SHA in this paper) in their book[18]:%
\textit{\ \textquotedblleft we shall make the unfounded assumption that the
molecules are randomly distributed throughout the volume and the number of
molecules lying near a particular molecule in a concentric shell ranging
from }$r$ \textit{to} $r+dr$ is $dN=N/V4\pi r^{2}dr$, \textit{with} $r\geq
\sigma .^{\prime \prime }$They derived the VDW equation of state from their
EHSE SHA, but the total potential energy and the partition function are
independent of the mean distance of molecules in their theory.

Now we put forth a new Space Homogeneity Assumption (SHA): \textit{In a
classical system in equilibrium with volume }$V$\textit{, particle number }$%
N $\textit{, when gravity and boundary effects are omitted, each particle
averagely occupies volume }$V/N$. We will demonstrate in detail later that
the SHA will exhibit the correlation between the mean distance of molecules
and the partition function.

Both the SHA here and the EHSE SHA should be regarded as independent
theoretical assumptions, which cannot be proved by statistical mechanics.
Their justifications lie in the agreement between results derived from them
and the experimental facts. Now we reevaluate the partition function under
the SHA.

We assume that the potential energy $u(r_{ij})$ between particle $i$ and
particle $j$ is shared by these two particles together and is divided
between both equally. We call $u_{i}=\sum\limits_{j\neq i}^{N-1}{\frac{1}{2}%
u(r_{ij})}$ the potential belonged to particle $i$ if only the pair
interaction exits. According to the SHA, the volume occupied by particle $i$
is equal to $V/N=1/n=\overline{r}^{3},$where $n$ is the particle number
density and $\overline{r}$ is the mean distance of molecules. For
convenience, we presume that this volume is a sphere with diameter $d=(6/\pi
n)^{1/3}=\alpha \overline{r}$ in a three-dimensional space, and particle $i$
is at its center. We then draw many spherical shells centered by particle $i$
outside this sphere. The thickness of each spherical shell is $d$. From
inside to outside, we number them $S_{1}$,$S_{2}$ and so on in turn. Then
the volume of the shell $S_{k}$ is $V_{k}=\frac{\pi }{6}\left[ {\left( {2k+1}%
\right) ^{3}-\left( {2k-1}\right) ^{3}}\right] d^{3}$ and the number of the
particles in this shell is $N_{k}=nV_{k}$. The potential energy of particle $%
i$ is $U_{i}=1/2\sum\limits_{k}N_{k}u(r_{k})$, where $u(r_{k})$ is the
potential contribution of a particle in the shell $S_{k}$ to particle $i$, $%
(k-1/2)d\leq r_{k}\leq (k+1/2)d$ and $\sum\limits_{k}N_{k}=N-1$. The total
potential of the system is $U=NU_{i}.$ The potential energy $u(r_{k})$ can
be expanded at the location $r_{k}=kd,$ which is the distance between the
midpoint of the shell $S_{k}$ and particle $i$: $u(r_{k})=u(kd)+u^{\prime
}(kd)(r_{k}-kd)+\cdot \cdot \cdot $. If only taking the first term of this
expansion, $i.e.$ $u(r_{k})\approx u(kd)$, we have the partition function%
\begin{equation}
Z_{0}=\frac{1}{N!\lambda ^{3N}}\exp \left( {-\beta \frac{N}{2}%
\sum\limits_{k}N_{k}{u(kd)}}\right) \int {dq,}  \label{1}
\end{equation}%
with $\lambda $ thermal wavelength. The Eq.(1) clearly indicates that the
potential (as does the partition function) of the system is correlative with
the mean distance. Such a relation plays a vital role in the discussion of
the phenomenon of liquid-gas coexistence. When in Eq.(1) taking the
Sutherland potential: $u(r)=\infty $ if $r<\sigma $, and $u(r)=-\varepsilon
(\sigma /r)^{3}$ if $r>\sigma $, where $\sigma $ is the diameter of the hard
core and $\varepsilon $ is a constant, we obtain the partition function of
VDW fluid: 
\begin{equation}
Z_{0}=\frac{1}{N!\lambda ^{3N}}\exp (\beta aN^{2}/V)(V-Nb)^{N},  \label{2}
\end{equation}%
with $a=\sum\limits_{k}{N_{k}\varepsilon \frac{\pi \sigma ^{3}}{12k^{3}}}$, $%
b=\frac{2}{3}\pi \sigma $. The VDW equation is $P=k_{B}T\frac{\partial }{%
\partial N}\left( {\ln Z_{0}}\right) _{T,N}$. The chemical potential of VDW
fluid can be obtained by

\begin{eqnarray}
\mu &=&-k_{B}T\left[ \frac{\partial \ln Z}{\partial N}\right]
_{V,T}=-2an+k_{B}T\ln \left[ \frac{n}{1-nb}\right]  \notag \\
&&+\frac{nk_{B}Tb}{1-nb}-\frac{3k_{B}T}{2}\ln \frac{2\pi mk_{B}T}{h^{2}}
\end{eqnarray}%
Letting 
\begin{equation}
\overline{\mu }=\frac{\mu }{k_{B}T}+\ln b+\frac{3}{2}\ln \frac{16\pi ma}{%
27bh^{2}},  \label{4}
\end{equation}%
and substituting Eq. (3) into (4), we have 
\begin{equation}
\overline{\mu }\left( T^{\ast },n^{\ast }\right) =\ln \frac{n^{\ast }}{%
3-n^{\ast }}-\frac{3}{2}\ln T^{\ast }+\frac{n^{\ast }}{3-n^{\ast }}-\frac{%
9n^{\ast }}{4T^{\ast }},  \label{5}
\end{equation}%
with reduced quantities $T^{\ast }=$ $T/T_{c}$ and $n^{\ast }=n/n_{c}$. The
Eq.(5) is a function of reduced quantities only. Now we can define the
reduced chemical potential $\overline{\mu }^{\ast }=\overline{\mu }/%
\overline{\mu }_{c}$. When $T<T_{c}$, there are three numerical values of $n$
corresponding to a given chemical potential $\mu $, which forecasts the
coexistence of multiple phases.

For a coexistent liquid-gas system we have the conditions of equilibrium: $%
P\left( {T,n_{g}}\right) =P\left( {T,n_{l}}\right) $ and $\mu \left( {T,n_{g}%
}\right) =\mu \left( {T,n_{l}}\right) $. Thus the solutions of these two
equations, $n_{g}=n_{g}\left( T\right) $ and $n_{l}=n_{l}\left( T\right) $
(the densities of the coexistent vapor and liquid, respectively), can be
obtained by a numerical method, as shown (dashed line) in Figure 1A. This
curve is equivalent to that given by Maxwell rule in thermodynamics.

\FRAME{ftbpFU}{4.7314in}{1.996in}{0pt}{\Qcb{Comparison between the
experiment and the theoretical results in this paper: A) Densities of
coexisting vapor and liquid. B) Vapor pressure. C) Latent heat. solid
line--experimental curve from Ref.[20-22], dashed line---VDW fluid,
dot-line---modified VDW.}}{}{3tu.eps}{\special{language "Scientific
Word";type "GRAPHIC";display "USEDEF";valid_file "F";width 4.7314in;height
1.996in;depth 0pt;original-width 6.8753in;original-height 2.1223in;cropleft
"0";croptop "1";cropright "1";cropbottom "0";filename
'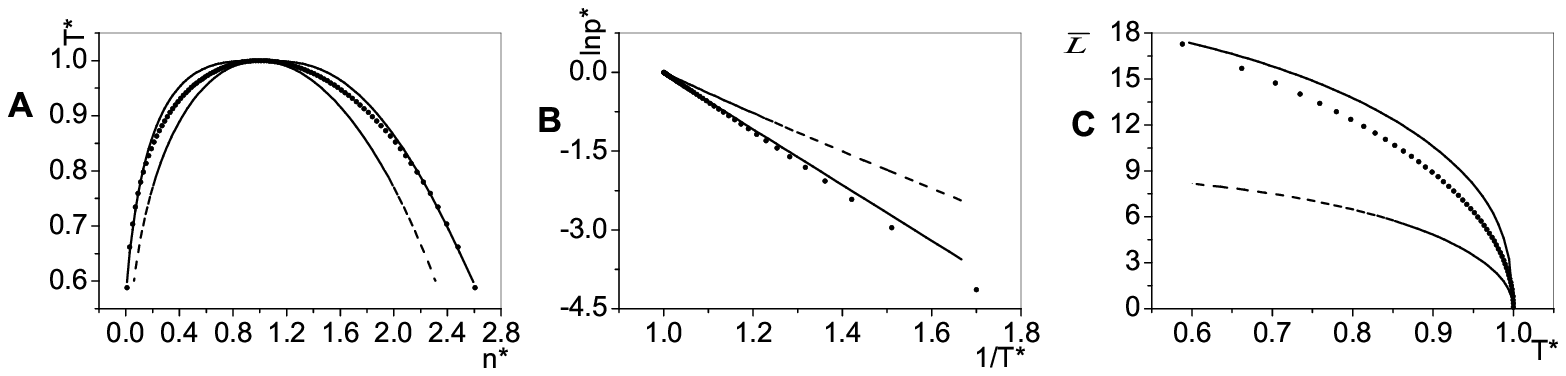';file-properties "XNPEU";}}

\bigskip

The dashed curve for VDW fluid in Fig.1A can be described by the empirical
formulae (firstly introduced by Wei Wang ) 
\begin{eqnarray}
n_{l}^{\ast } &=&1+0.497\left( 1-T^{\ast }\right) ^{1.086}  \notag \\
&&-0.56\left( 1-T^{\ast }\right) ^{1.53}+2\left( 1-T^{\ast }\right) ^{0.5},
\end{eqnarray}%
\begin{eqnarray}
n_{g}^{\ast } &=&1+0.497\left( 1-T^{\ast }\right) ^{1.086}  \notag \\
&&+0.56\left( 1-T^{\ast }\right) ^{1.53}-2\left( 1-T^{\ast }\right) ^{0.5}.
\end{eqnarray}%
or 
\begin{equation}
\frac{n_{l}^{\ast }-n_{g}^{\ast }}{2}=-0.56\left( 1-T^{\ast }\right)
^{1.53}+2\left( 1-T^{\ast }\right) ^{0.5},  \label{8}
\end{equation}%
\begin{equation}
\frac{n_{l}^{\ast }+n_{g}^{\ast }}{2}=1+0.497\left( 1-T^{\ast }\right)
^{1.086}.  \label{9}
\end{equation}

Formula (8) suggests the critical exponent $\beta =0.5$ when $T^{\ast
}\rightarrow 1$. Formula (9) states the well-known law of the rectilinear
diameter. Comparing with the numerical values of the solutions , the
inaccuracy of the formula (6) is less than $0.1\%$ when $T^{\ast }>0.6$, and
the inaccuracy of the formula (7) is less than $0.1\%$ when $T^{\ast }>0.71$%
. The percentage of inaccuracy increases as the temperature drops below $%
T^{\ast }=0.71$.

Substituting them into VDW equation, we then get the vapor pressure $P=P(T)$%
. The reduced vapor pressure is expressed as 
\begin{equation}
P^{\ast }=\frac{8n_{g}^{\ast }T^{\ast }}{3-n_{g}^{\ast }}-3n_{g}^{\ast 2}=%
\frac{8n_{l}^{\ast }T^{\ast }}{3-n_{l}^{\ast }}-3n_{l}^{\ast 2}.  \label{10}
\end{equation}%
as shown in Figure 1B (dashed line). Substituting $\mu ^{\ast }$ into Eq.
(10), we obtain another expression of vapor pressure 
\begin{equation}
P^{\ast }=\frac{3n_{g}^{\ast 2}\exp \left[ \frac{9}{4T^{\ast }}\left(
n_{l}^{\ast }-n_{g}^{\ast }\right) -\frac{3}{8T^{\ast }}\left( n_{l}^{\ast
2}-n_{g}^{\ast 2}\right) \right] -n_{l}^{\ast 2}}{1-\exp \left[ \frac{9}{%
4T^{\ast }}\left( n_{l}^{\ast }-n_{g}^{\ast }\right) -\frac{3}{8T^{\ast }}%
\left( n_{l}^{\ast 2}-n_{g}^{\ast 2}\right) \right] }.  \label{11}
\end{equation}%
This relation can be expressed by the empirical formula

\begin{equation}
\ln P^{\ast }=3.5204-3.5660/T^{\ast }  \label{12}
\end{equation}

There is a gap of internal energies between the two phases. From Eq.(2) the
potential energies per particle in vapor and liquid are $U_{ig}=-an_{g}=-1/2%
\sum\limits_{k}{N_{k}u}(k\alpha \overline{r_{g}})$ and $U_{il}=-an_{l}=-1/2%
\sum\limits_{k}{N_{k}u}(k\alpha \overline{r_{l}})$ respectively. The
difference between the two is $dU_{gl}=U_{ig}-U_{il}=a(n_{l}-n_{g})$, which
is the gap of internal energies between the two phases due to the different
mean distances, as shown in Fig.2, where $U^{\ast }=U/U_{c}$. The average
distance of gas molecules is larger than that of liquid molecules. The
interactions between gas molecules and the interactions between liquid
molecules can be described by the same interaction potential, such as
Sutherland potential. The difference between the mean distances of molecules
in the two phases results in the difference of molecular potential energies
in the two phases. When a liquid molecule vapors and becomes a gas molecule,
there is an interaction potential energy gap between the molecules in the
two phases $dU_{gl}$. The gap exists only under the SHA and the mean
distance expansion, in contrast with Mayer's theory in which there was no
gap. In fact, the partition function (2) was already formally presented in
some of the previous works, but it is with different meanings and the
comparable inferences on phase transition could not be arrived at as no gap
existed in their discussion.

\FRAME{ftbpFU}{3.1773in}{2.0565in}{0pt}{\Qcb{ The relation between the
potential energy per particle and densities. The curve $a-g-l-d$ is an
isotherm of pressure $P^{\ast }$. When pressure runs from $a$ to $g$, or
from $g-l$ the corresponding potential energy changes from U$_{i}^{\ast }$(n$%
_{a}^{\ast }$) to U$_{i}^{\ast }$(n$_{g}^{\ast }$). The gap $U_{il}^{\ast
}-U_{ig}^{\ast }$ is shown.}}{}{up.eps}{\special{language "Scientific
Word";type "GRAPHIC";display "USEDEF";valid_file "F";width 3.1773in;height
2.0565in;depth 0pt;original-width 2.8323in;original-height 2.3281in;cropleft
"0";croptop "1";cropright "1";cropbottom "0";filename
'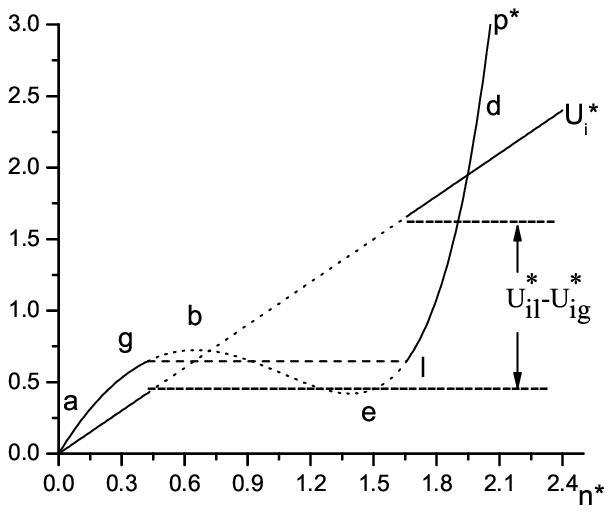';file-properties "XNPEU";}}

\bigskip

The molar latent heat can be expressed simply by

\begin{eqnarray}
L &=&N_{A}\left( {dU_{gl}+Pdv_{gl}}\right)  \notag \\
&=&N_{A}\left[ {a\left( {n_{l}-n_{g}}\right) +P\left( {1/n_{g}-1/n_{l}}%
\right) }\right] ,  \label{13}
\end{eqnarray}%
which is equivalent to the Clapeyron's equation[19] and $N_{A}$ is
Avogadro's constant. The meaning of this formula is very clear: the mean
distances between molecules are different when the system is in different
phases. This difference leads to the gap of the intermolecular potential
between the two phases, which together with the work done due to volume
difference between two phases engenders latent heat. Other theories cannot
give latent heat also. We define $\overline{L}=L/P_{c}v_{c}$ and $\overline{L%
}$\ can be written as

\begin{equation}
\overline{L}=3(n_{l}^{\ast }-n_{g}^{\ast })+P^{\ast }(\frac{1}{n_{g}^{\ast }}%
-\frac{1}{n_{l}^{\ast }}).  \label{14}
\end{equation}%
which is a function of reduced quantities $P^{\ast }(T^{\ast })$\ and $%
n^{\ast }(T^{\ast })$\ only. Fig.1C shows the relation between latent heat\ $%
\overline{L}$ and temperature $T^{\ast }$. The solid-curve is drawn from
experimental data of Argon[22]. The dash-curve is the latent heat of VDW
fluid from Eq.(14). The consistencies between the latent heat calculated by
the Eq.(13) and that by the Clapeyron's equation provide a strong support to
the new theory in this letter.

From Figure1 we find that the densities of coexistent vapor-liquid, the
vapor pressure and the latent heat of VDW fluid are qualitatively correct,
but quantitatively incorrect, when compared with the experimental
data[20-22]. This inconsistency is due to the incorrect potential we have
chosen.

Using the effective potential $u(r)=-\varepsilon T^{S}(\sigma /r)^{M}$
(referred to as modified VDW fluid) with S=-0.49 and M=2.83 (in Mayer's and
Eyring's works, we should have $M\geq 3$ to keep convergence, but there is
no such restriction in mean distance expansion), instead of the attractive
potential $u(r)$ in Sutherland potential, and following the above standard
calculation procedure, we obtain the better results(see dot-line in Fig.1).

Now let us review what is new in this work. The starting point of this work
is the SHA. The difference between the SHA and the previous works, such as
Mayer's and Eyring's work, is the following: a). The EHSE SHA claimed that
the number of the particles in a volume $d^{3}r$ is $dN=N/V4\pi r^{2}dr$
whereas the SHA postulates that each particle occupies a volume $V/N$. b).
Mayer and Eyring assumed that the distributions of particles are
nonexclusive, i.e. the distribution of an arbitrary particle is not affected
by other particles. On the contrary, the SHA postulates that the
distributions of particles are exclusive, i.e. each volume element $V/N$
can, but can only lodge one particle averagely. This is a fundamental
difference between mean distance expansion and Mayer's cluster expansion.
c). The intermolecular interactions in Mayer's and Eyring's calculations are
independent of the mean distance $\overline{r}$, meaning that when the mean
distance or the density varies, the intermolecular interactions do not
change. As a result, based on its interactions, we cannot distinguish which
phase the system is in: vapor or water. But in this work, when the density
of the system goes from $n_{g}$ to $n_{l}$, or from $n_{a}$ to $n_{g},$the
interaction will change as the mean distance varies.

The new idea in this work is that the total intermolecular potential energy
and partition function for a classical system in equilibrium are relative to
the average distance $\overline{r}$ of molecules (or density). Starting from
it, we obtain the energy gap between two phases, and a new explanation for
the first order gas-liquid phase transition is given as follows: first order
gas-liquid phase transition does not correspond to singularity of
thermodynamic functions. In Fig.2 if a system runs from $a-g$ on the
isotherm $a-g-l-d,$ the corresponding potential energy increases as density
changes. At $g$ or $l$, there are two possible phases wiith different
intermolecular potential. These different fluid phases correspond to
different solutions of Eqs. $P\left( {T,n_{g}}\right) =P\left( {T,n_{l}}%
\right) $ and $\mu \left( {T,n_{g}}\right) =\mu \left( {T,n_{l}}\right) $.,
which form two branches of the solutions. For each branch $\mu _{l}\left( {%
T,n_{l}}\right) $ or $\mu _{g}\left( {T,n_{g}}\right) $, either function $%
\mu _{l}\left( {T,n_{l}}\right) $ or $\mu _{g}\left( {T,n_{g}}\right) $ are
analytic (see ref.[19] for detail). When $T=T_{c},$ we have $%
n_{l}=n_{g}=n_{c}$ and the gap $dU_{gl}\rightarrow 0$. Here there is no
request for the thermodynamic limit in strict mathematical sense.

\begin{acknowledgement}
I would like to express my thanks to Prof. L.Yu and Prof. Z.R.Yang for their
helpful conversations and to my graduate students J.X.Tian, W.Wang,
S.H.Zhang and Y.Lyu for their enthusiastic participation. This work was
supported by the NSFC under No.10275008.
\end{acknowledgement}

\bigskip \textbf{References }[1]. J.V.Sengers, Ann.Rev. Phys.Chem. 37,
189(1986).

[2]. Y. Levin, Phys.Rev.Lett. 83, 1159 (1999).

[3]. J.V.Sengers et al., Ed., Equations of State for Fluids and Fluid
Mixtures (Elsevier, Amsterdam, 2000)..\ \ \ \ \ \ \ \ \ \ \ \ 

[4]. L.Yu, B.L.Hao, Phase Transition and Critical Phenomena, (Science Press,
Beijing, ed.2, 1992, in Chinese)

[5]. S.J. Lee, A.Z. Mekjian, Phys.Rev. C68, 014608 (2003).

[6]. O. Tchernyshyov, S. L. Sondhi, Nucl.Phys. B639, 429-449 (2002).

[7]. N.A. Lima, M.F. Silva, L.N. Oliveira, K.Capelle, Phys. Rev. Lett. 90,
146402 (2003).

[8]. D. Pini, G.. Stell, N.B. Wilding, J. Chem. Phys. 115, 2702 (2001)

[9]. K. A. Bugaev, M.I. Gorenstein, I.N. Mishustin, W. Greiner, Phys.Rev.
C62, 044320(2000).

[10]. A. Gammal, T. Frederico, Lauro Tomio, Ph. Chomaz, Phys. Rev. A 61,
051602(R) \ \ \ \ \ \ \ \ \ \ \ \ \ \ (2000)..

[11]. J.X.Tian, Y.X.Gui, J.Phase Equili. 24, 533 (2003).

[12]. J.M.H.Levelt Sengers, Physica 73, 73(1974).

[13]. Y.G. Ma, Phys.Rev.Lett. 83, 3617(1999).

[14]. J. E. Santos, U. C. Tauber, Eur. Phys. J. B 28, 423 (2002).

[15]. S. Prestipino, P. V. Giaquinta, J. Phys.: Condens. Matter 15, 3931
(2003).

[16]. J.X.Tian, Y.X.Gui, J.Phase Equili. 24, 533 (2003).

[17].Supercritical Fluids, NATO, Kemer, Antalya, Turkey, July 18-31, 1993,
E.Kiran and \ \ \ \ \ \ \ \ \ \ \ J.M.H.Levelt Sengers Eds. (Kluwer
Academic, Dordrecht, 1994)

[18] H. Eyring, D. Henderson, B. J. Stover, and E.M. Eyring,

Statistical Mechanics and Dynamics, (John Wiley and Sons, New

York, ed.2, 1978) pp. 620-21.

[19] Y.X.Gui, Gas-Liquid Transition In Statistical
Mechanics(cond-mat/0404665),

[20] E.A.Guggenheim, J.Chem.Phys, Vol.13, Num.7 (1945).

[21] E.A.Guggenheim, thermodynamics, (Oxford Univ. Press, New York,

1966), p.37-38.

[22] S.Angus, B.Armstrong, Eds., International Thermodynamic Tables

of the Fluid State, Argon, 1971, (Butterwo, London, 1972).

[23] D.J.Berthelot, J.Phys. 8, 263 (1899).

\end{document}